\documentclass[conference,a4paper]{IEEEtran}
\usepackage{amsmath,amsthm,amssymb,amsfonts}
\usepackage{color}
\usepackage{graphicx}
\usepackage{subfigure}
\usepackage{dsfont}
\usepackage{cite}
\usepackage{caption}
\usepackage{hyperref}
\usepackage{algorithm}
\theoremstyle{definition}

\theoremstyle{remark}

\graphicspath{{pics/}}
\def\BibTeX{{\rm B\kern-.05em{\sc i\kern-.025em b}\kern-.08em
    T\kern-.1667em\lower.7ex\hbox{E}\kern-.125emX}}
\newcommand{\orcid}[1]{\href{https://orcid.org/#1}{ORCID: #1}}    
\begin{document}
\title{Early-Stopped Technique for BCH Decoding Algorithm Under Tolerant Fault Probability}
\author{%
   \IEEEauthorblockA{%
    Hong-fu Chou\\
    Interdisciplinary Centre for Security, Reliability, and Trust, University of Luxembourg\\
    Email: hungpu.chou@uni.lu\\
     \textsuperscript{\orcid{0000-0001-7932-6297}}}
 }%



\maketitle

\begin{abstract}
 In this paper, a technique for the Berlekamp-Massey(BM) algorithm is provided to reduce the latency of decoding and save decoding power by early termination or early-stopped checking. We investigate the consecutive zero discrepancies during the decoding iteration and decide to early stop the decoding process. This technique is subject to decoding failure in exchange for the decoding latency. We analyze our proposed technique by considering the weight distribution of BCH code and estimating the bounds of undetected error probability as the event of enormous stop checking. The proposed method is effective in numerical results and the probability of decoding failure is lower than $10^{-119}$ for decoding 16383 code length of BCH codes. Furthermore, the complexity compared the conventional early termination method with the proposed approach for decoding the long BCH code. The proposed approach reduces the complexity of the conventional approach by up to 80\%. As a result, the FPGA testing on a USB device validates the reliability of the proposed method.
\end{abstract}
\section{Introduction}
Flash memory \cite{Seii}\cite{Yoshio} performs as the main non-volatile storage device, and the flash interface unit is applied for system-on-chip (SoC) products. The market size of NAND flash memories is still growing and is projected to see a compound annual growth rate of 6.39\%\cite{NANDmarket}. Flash memory provides a low-power solution for storage systems and, small size and the light form factor are the essential properties for this type of storage. The flash interface unit\cite{Wei} provides basic flash commands which can be used by the main central processing unit(CPU) to access data from the flash memory. It is assumed that the flash memory is non-removable since the flash memory is used to initiate the boot process based on information from the firmware. 

Flash memory plays an important role in the storage device to execute the tasks to be performed by the main CPU. The tasks are literally to read and write files and are identical to any generic file system. The flash interface unit has mainly provided a reliable component for graphics and multimedia processors and has been applied to digital televisions, car navigation systems, and mobile applications. To support multimedia applications, flash interface units have been optimized for large block read and write, as presented in \cite{Yu}. To minimize the main CPU interaction, the flash interface unit supports direct memory access (DMA)\cite{Rota} when transferring from the flash memory to the system DRAM memory. 

In the SoC applications, all of the boot information is generally stored in flash memory. The flash memory includes a number of partitions for the boot loader code and the flash file system are created in the flash memory. In \cite{Rota}, the DMA interacts with the error control coding (ECC) block, which provides two main purposes. The first is to generate the ECC bytes and program in the spare area, and the second is to correct the data in the data buffer. Consequently, the ECC engine is a critical issue regarding system performance. The chip area is dominated by the ECC decoder, comprising a high percentage of the flash controller. 

The Bose-Chaudhuri-Hocquenghem (BCH) code has become the ultimate solution for the ECC engine in recent years. In coding theory, the BCH codes form a class of cyclic error-correcting codes that are constructed using finite fields. The decoding algorithm is based on a feasible implementation where the Berlekamp-Massey (BM) algorithm \cite{Berklekamp} has been widely selected in typical examples. The complexity of the decoding is competitive with respect to the BM properties of the linear feedback shift register. However, system latency suffers from larger $t$ error correction capability which requires $2t$ iterations of conventional BM decoding and common applications require high error-correcting capability. The long decoding time has become a bottleneck in the system performance while using BM decoding. The error distribution for flash memory shows that few errors at the beginning of its usage and the low number of errors dominate the majority of the probability that will occur within a code block. In order to overcome this degradation, early termination of BM decoding is necessary to improve the system performance for high-speed applications. In \cite{Liu}, the authors adopt a restricted Gaussian elimination on the Hankel structured augmented syndrome matrix to reinterpret an early-stopped version of the Berlekamp-Massey algorithm. This approach has proven the minimal iterations $t+e$ of the Berlekamp-Massey algorithm where $e$ is the number of error bits. Following the thread of \cite{CLChen}, the author presents a feasible approach for early termination but the investigation of malfunction probability was present in \cite{Sanvate}. 

In this paper, the probability of decoding failure is considered in exchange for early-stopped BM decoding feasibility. The proposed technique terminates conventional BM decoding after less than $t+e$ iterations so as to reduce redundant latency. However, the proposed technique is subject to the decoding failure problem. The probability that a detection error will occur must be evaluated to ensure the reliability of the proposed approach. Consequently, we propose an early-stopped technique for BM decoding by observing certain conditions while performing decoding iterations. In Section II, we present the early-stopped checking procedure of BM decoding by observing consecutive zero discrepancies. Since zero discrepancies provide the information of detectable decoding, it is an interesting problem to estimate the undetectable decoding after consecutive zero discrepancies. We provide an estimation of the enormous early-stopped checking by means of the probability of undetected error probability in \cite{Kim}. After combining the early-stopped checking criterion in \cite{CLChen}, we propose our approach. In Section III, the complexity analysis is presented to compare with the conventional early-stopped BM approach. In Section IV, the numerical results are presented to evaluate the feasibility of a practical application. Conclusions are presented in Section V.

\section{Early stopped approach based on the view of discrepancy for the BM algorithm}
In coding theory, BCH codes \cite{LinCostello}\cite{Peterson}  are constructed using polynomials over a finite field (also called the Galois field and is denoted as GF(q)). One of the key features of BCH codes is that, during code design, there is precise control over the number of symbol errors that are correctable by the code. In particular, it is possible to design binary BCH codes that can correct multiple-bit errors in discrete distribution under a correction capability of $t$ bits. Another advantage of BCH codes is the ease with which they can be decoded, namely, via an algebraic method known as syndrome decoding. This simplifies the design of the decoder for these codes, using small low-power electronic hardware.

BCH codes are used in applications such as satellite communications, compact disc players, DVDs, disk drives, solid-state drives, etc.

There are many algorithms for decoding BCH codes. The most common follow this general outline:
\begin{description}
\item[1.] Calculate the syndromes for the received vector
\item[2.] Determine the number of errors $v$ and the error locator polynomial $N(x)$ from the syndromes
\item[3.] Calculate the roots of the error location polynomial to determine the error locations $X_{i}$
\item[4.] Calculate the error values at those error locations
\item[5.] Correct the errors
\end{description}
The decoding algorithm may determine that the received vector contains too many errors and cannot be corrected. For example, if the number of errors is greater than the correction capability, then the correction would fail. In a truncated (not primitive) code, an error location may be out of range. If the received vector has more errors than the code can correct, the decoder may unknowingly produce an apparently valid message that is not the one that was sent.

In order to determine any possible solutions to shorten the BM decoding process, based on the result in \cite{CLChen} and \cite{Liu}, we classify the solutions in two conditions as follows.
\begin{description}
\item[$Condition~1$:]\hspace{0.1cm}\\
For the $u$-th iteration of the BM algorithm, the discrepancy at iteration $u$ is presented as $d_{u}$, and any discrepancies in the next t-$l_{u}$-1 steps of the iteration are zero.
\item[$Condition~2$:]\hspace{0.1cm}\\
If the number of errors in the received polynomials is $v$, only $t+v$ steps of the iteration are needed in order to determine the error-location polynomials.
\end{description}
\subsection{Heuristics for consecutive zero discrepancies }
Following the thread of $Condition~2$, the probability of the enormous event based on the view of the discrepancy is investigated as follows. The discrepancies in certain iterations equal to zero, as shown in $Condition~1$ represent the detection capability reach in a certain level of $l_{u}$ iterations, i.e. $l_{u}=v$, where $v$ is the number of error bits hypothesized by our proposed approach.\\
$Heuristic~1$:\hspace{0.1cm} Let a BCH code $\zeta$ have minimum Hamming distance $d\geq 2t+1$ and consider that $\zeta^{v+\kappa} \subset \zeta$ denotes a BCH code subset with minimum Hamming distance $d_s\geq v+\kappa$ and $\kappa$ is the number of consecutive zero discrepancies for the $v$-th iteration of BM algorithm. The next $\kappa$ steps actually occurred with $v+\kappa\leq 2t $. $Rationale$: The Hamming distance for the received codeword $r$ and the transmitted codeword $c$ is presented as $d(r,c)=i$, $i<t$, where $c\in\zeta^{v+\kappa}$. \\
$Heuristic~2$:\hspace{0.1cm}The error pattern $\xi$ defects the codeword $c$, it can also be presented as $r=c+\xi$ and $d(r,c)=d(\xi,c)$. $Rationale$: Assume $e=v$ and $e$ denotes the exact number of error bits caused by the channel without the decoding fault. Otherwise, a malfunction occurs when the location of the error pattern is beyond the detection capability at $l_{u}=v+\kappa$ iteration which indicates the case of $v+\kappa<e$. 
\subsection{Numerical Analysis of fault probability for the proposed early stopped technique}
Based on the above heuristics, the error event of observing consecutive zero discrepancies during decoding iterations is invested as follows. A non-zero discrepancy occurs after performing $v+\kappa$ BM decoding iterations and the codeword $c \in \left\{\zeta^{v+\kappa}-\zeta\right\}$ which results in the proposed technique failing to provide a correct BM decoding. Hence, the probability of malfunction is given as follows.
\begin{align*}
P_{mf}=p[v+\kappa<e]~~~~~~~~~~~~~~~~~~~~~~~~~~~~~~~~~~~~~~~~~~~~~~\\
=\sum^{t}_{i=0}P[d(r,c)=i|c\in\zeta^{i+\kappa}-\zeta]~~~~~~~~~~~~~~~~~~~~~~~~~~~~~\\
=\sum^{t}_{i=0}P[d(\xi,c)=i|c\in\zeta^{i+\kappa}]-P[d(\xi,c)=i|c\in\zeta]~~~~(1)\\
\end{align*}
According to \cite{Kim}, the bounds of the probability $P_{ud}$ that an undetected error will occur can be bound by the assumption of a long codeword length $n$ and $m$ is equal to the message length, 
\begin{align*}
P_{ud}=\sum^{t}_{i=0}P[d(\xi,c)=i|c\in\zeta]~~~~~~~~~~~~~~~~~~~~~~~~~~~\\
\cong 2^{-mt}\sum^{t}_{s=0} \binom{n}{s}\sum^{n}_{h=t+1}\binom{n}{h}\varepsilon^{h}(1-\varepsilon)^{n-h}~~~~~~~~~~(2)\\
\end{align*}
The undetected error probability of the difference between upper and lower bounds is limited to 1\%. We further extend the bounds of the probability of an error pattern given by \cite{Meera} and \cite{Kim}. The conditional probability of a BCH code $\zeta^{d'}$ that has minimum Hamming distance $d'$ is interpreted as follows. 
\begin{align*}
P[d(\xi,c)=i|c\in\zeta^{d'}]\cong\sum^{n}_{h=(d'+1)/2}\binom{n}{h}\varepsilon^{h}(1-\varepsilon)^{n-h}~~~~~~(3) \\
\end{align*}
Substituting (3) into (1), the probability of malfunction can be estimated as
\begin{align*}
P_{mf}\cong 2^{-mt}[\sum^{t}_{s=0}\binom{n}{s}\sum^{n}_{h=(s+\kappa+1)/2}\binom{n}{h}\varepsilon^{h}(1-\varepsilon)^{n-h}\\
-\sum^{t}_{s=0}\binom{n}{s}\sum^{n}_{h=t+1}\binom{n}{h}\varepsilon^{h}(1-\varepsilon)^{n-h}]~~~~~~(4) \\
\end{align*}
Furthermore, (4) can be simplified further by bounds of the type considered in \cite{Kim} and define $\lambda _{1}=(s+\kappa+1)/(2n)$ and  $\lambda _{2}= (t+1)/n$.  
\begin{align*}
P_{mf}\cong 2^{-mt}\sum^{t}_{s=0}\binom{n}{s} [2^{-nE(\lambda _{1}, \varepsilon )}- 2^{-nE(\lambda _{2}, \varepsilon )}]~~~~~(5) \\
\end{align*}
where $E(\lambda, \varepsilon)$ is the relative entropy between the binary probability distribution $\lambda$ and $\varepsilon$.
\begin{align*}
E(\lambda, \varepsilon)=H(\varepsilon)+(\lambda - \varepsilon)H(\varepsilon)-H(\lambda)~~~~~~~~~~~~~~(6)\\
=\lambda log_{2}(\lambda/\varepsilon)+(1-\lambda)log_{2}((1-\lambda)/(1-\varepsilon)) ~~~~~~~~~~~~
\end{align*}
Based on the above observing $d_{j}$ discrepancies during BM iteration, we illustrate the proposed early-stopped checking method, which is described below. The proposed method is denoted as the early-stopped(ES) version, and we provide three different versions. For BM decoding of the $j$-th iteration, we observe the following discrepancy based on the proposed method.
We denote that $\delta_{max}$ represents the maximum error location degree of the BM algorithm.
\begin{algorithm}
\caption{ES version 1}\label{alg:cap}
Beginning from $j=4$ as $j$-th iteration of the BM algorithm, verify the following steps:\\
1. Check Case A: $t$+ $\delta_{max}/2$ = $j$ \\
2. Check Case B: $d_{j}$, $d_{j-1}$, $d_{j-2}$ and $d_{j-3}$ are all zero.\\
3.  If Case A and Case B are satisfied, terminate the BM decoding. Otherwise, proceed to the next BM iteration and return to Step 1.
\end{algorithm}
\begin{algorithm}
\caption{ES version 2}\label{alg:cap}
Beginning from $j=6$ as $j$-th iteration of the BM algorithm, verify the following steps:\\
1. Check Case A: $t+ \delta_{max}/2$ = $j$\\
2. Check Case B: $d_{j}$, $d_{j-1}$, $d_{j-2}$, $d_{j-3}$, $d_{j-4}$, $d_{j-5}$ are all zero.\\
3. If Case A and Case B are satisfied, terminate the BM decoding. Otherwise, proceed to the next BM iteration and return to Step 1.
\end{algorithm}
\begin{algorithm}
\caption{ES version 3}\label{alg:cap}
Beginning from $j=\kappa$ as $j$-th iteration of the BM algorithm, verify the following steps:\\
1. Check the Case A: $d_{j}$, $d_{j-1}$, ..., $d_{j-\kappa+1}$ are all zero.\\
2. If Case A is satisfied, terminate the BM decoding. Otherwise, proceed to the next BM iteration and verify Step 1.\\
$\kappa$ is set to 4, 5 or 6 before simulation.
\end{algorithm}

ES version 1 in Algorithm 1 for checking 4 consecutive zero discrepancies and ES version 2 in Algorithm 2 for checking 6 zeros are presented to summarize a combination of early-stopping approaches considering \cite{CLChen} and our technique. However, ES version 3 in Algorithm 3 is the main core of our proposed approach to reveal the best complexity reduction. 

\section{Complexity analysis}
\begin{figure}[hbt]
\includegraphics[width=3.7in]{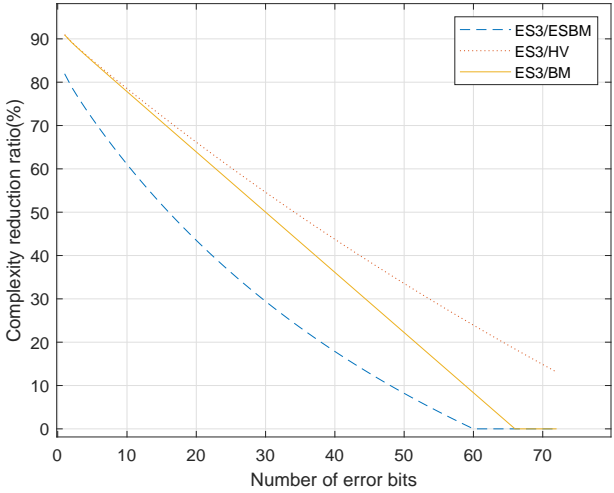}
\caption{The complexity reduction of the proposed technique with t=72.}
\label{complxesbm}
\end{figure}

The early stopped technique enjoys saving processing time and lowers power consumption. In this section, the analysis of multiplicative complexity is presented. Thanks to the author in \cite{Liu} that the upper bound of complexity analysis can be applied to evaluate the proposed technique by comparing it with the conventional BM algorithm and its related early-stopped technique. Since our proposed technique stops the conventional BM algorithm by certain conditions, the complexity of decoding can be computed by considering stopping the conventional BM algorithm at $e+\kappa$ iterations. Following the thread in \cite{Liu}, the multiplicative complexity $C_{ES3}$ of the proposed ES version 3 is upper bound by $2e(e+\kappa)-1$ which require at most $e+\kappa$ steps to check the discrepancies $d_{j}$. We summarize the comparison in Table I to show the merit of our proposed technique. $e$ denotes the exact number of error bits caused by the channel. To compare with the proposed technique, the conventional BM algorithm and conventional early-stopped technique enjoy low complexity when decoding the short codeword BCH code with a small t. However, the complexity of our proposed technique is not related to the parameter $t$ and is only dominated by $e^2+e\kappa$ which is quite beneficial for decoding long BCH code with larger correcting bits $t$. The complexity analysis results contribute to the applications such as NAND flash and future satellite communication. A 16384 code length BCH code with large $t=72$ is considered. For an example of $t=72$, $e=2$ and $\kappa=6$, $1-C_{ES3}/C_{ESBM}$ denote as the complexity reduction ratio of the proposed technique is equal to 79\%. We present the complexity reduction ratio in Fig.~\ref{complxesbm} and the proposed technique can reach up to 80\% improvement over the early-stopped approach in \cite{Liu}. The complexity reduction comes from taking the risk of decoding failure. Hence, we investigate the probability of decoding failure for the proposed technique in the following section.      
\begin{table}[ht]
\centering
\caption{COMPARISON OF UPPER BOUNDS OF FINITE-FIELD MULTIPLICATIVE COMPLEXITY }
\begin{tabular}[b]{|c|c|c|c|}
\hline
$C_{ESBM}$\cite{Liu}&$C_{HV}$&$C_{BM}$&$C_{ES3}$\\
\hline
$te+e^{2}-1$&$2te+\frac{1}{2}(e^{2}-e)$&$2et-1$&$2e(e+\kappa)-1$\\
\hline
\end{tabular}
\end{table}

\section{Numerical results}
 The proposed early stopped technique has the capability to reduce the decoding latency. For example, the case of t error correcting which is equal to 72 leads to a huge cost of the area to implement the BCH decoder and the decoding latency of BM decoding degrades the system performance of the DMA accessing the flash memory. The authors in \cite{Kim} obtained bounds on the probability of undetected errors in binary primitive BCH codes by applying the result to the code and showed that the bounds are quantified by the deviation factor of the true weight distribution from the binomial-like weight distribution. This approach presents a promising prediction for us to investigate that a long primitive BCH code can be robust to applying an early-stopped technique for a NAND flash system.

\begin{figure}[hbt]
\includegraphics[width=3.6in]{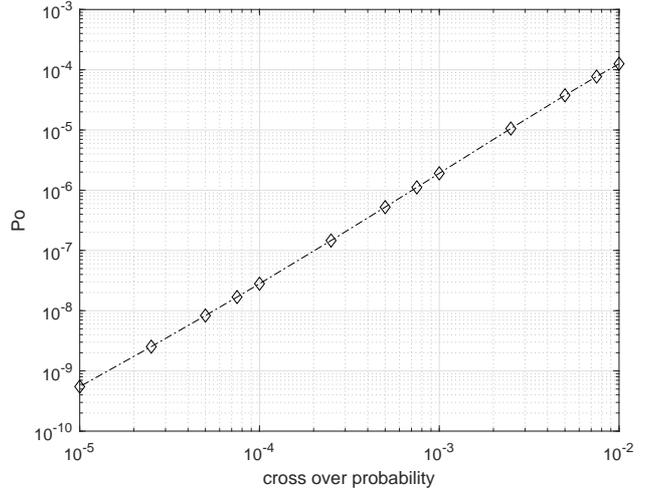}
\caption{The probability of enormous event during early termination checking using $GF(2^{5})$ BCH code t=3.}
\label{esbm1}
\end{figure}

\begin{figure}[hbt]
\includegraphics[width=3.7in]{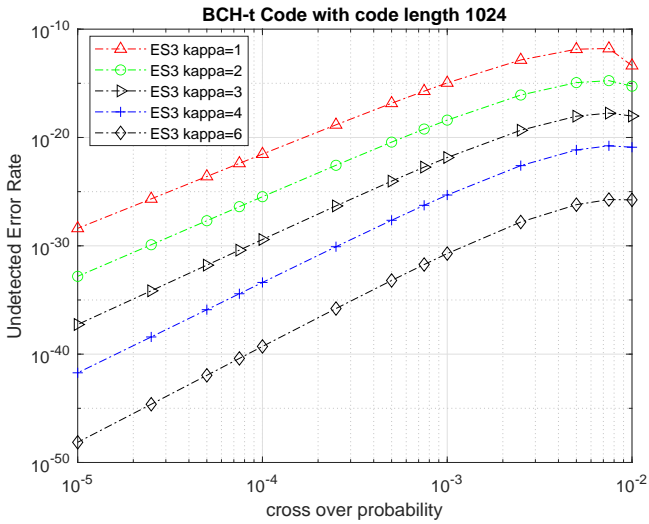}
\caption{The probability of undetected errors during early termination checking for ES version 3 using $GF(2^{10})$ BCH code t=17.}
\label{esbm3}
\end{figure}

First, we consider a BCH code with a length that is equal to 31 in $GF(2^{5})$, and that can correct $t=3$, which has an outcome of $2^{31}$ codewords. During the decoding of the received codewords used to compute the discrepancy, we consider the following case in Table II.
\begin{table}[ht]
\centering
\caption{A FAILURE CASE OF EARLY STOPPED CHECKING}
\begin{tabular}[b]{|c|c|c|c|c|}
\hline
Discrepancy&$>0$&0&0&d'\\
\hline
BM iteration&1&2&3&4\\
\hline
\end{tabular}
\end{table}
If we observe that the number of discrepancies is consecutively zero, we can compute the probability of a failure event occurring if d' is equal to non-zero. A conditional failure event can cause the proposed method to fail to decode a correct codeword which is subject to the observation of consecutively zero discrepancies. 
The failure rate Po is illustrated based on equation (4) for a certain degree of non-zero discrepancy during each iteration. In Fig~\ref{esbm1}, it can be observed that Po has the bound of $1.25562\times10^{-4}$. This simple example can figure out the problem causing the decoding failure.
Consequently, it is interesting to investigate how should we set the parameter $\kappa$. The probability of enormous early-stopped checking for the proposed ES version can be calculated using equation (5). In Fig~\ref{esbm3}, a BCH code with a length 1024 and $t=17$ is presented to show that the highest probability of an enormous event for proposed ES version 3 is $1.63752\times10^{-12}$ for $\kappa =1$, $1.7629\times10^{-15}$ for $\kappa =2$ and $1.77413\times10^{-18}$ for $\kappa =3$ respectively. As a result, we trade the failure probability with the early-stopped technique is not good enough while we use $\kappa =1,2,3$. In particular, a threshold of $\kappa$ is set as $\kappa \geqslant 4$ to obtain the result with the probability of an enormous event as $1.7005\times10^{-21}$ for $\kappa =4$ and $1.85605\times10^{-26}$ for $\kappa =6$.      

\begin{figure}[hbt]
\includegraphics[width=3.7in]{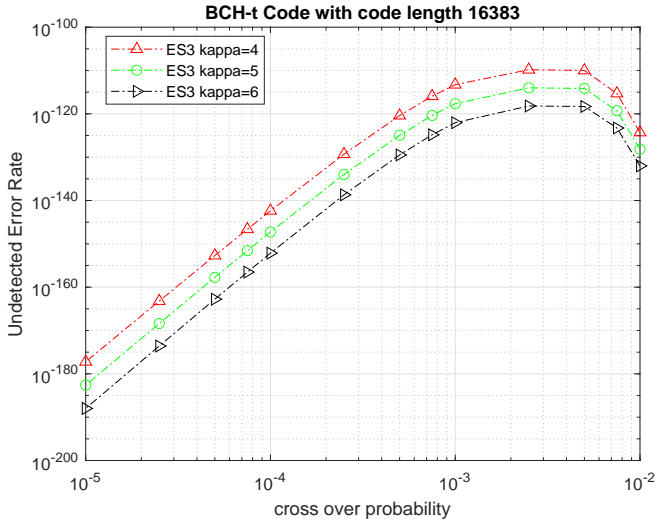}
\caption{The probability of undetected errors during early termination checking for ES version 3 using $GF(2^{14})$ BCH code t=72.}
\label{esbm2}
\end{figure}

\begin{figure}[hbt]
\includegraphics[width=3in]{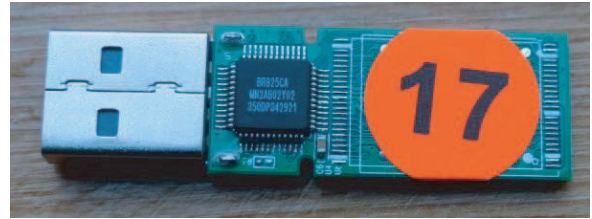}
\caption{A photo of the commercial product applying the proposed technique}
\label{825ca}
\end{figure}

Furthermore, we show that the problem of decoding failure caused by early-stopped technique can be neglected with the nature of long BCH codes. By using equation (5) as shown in Fig~\ref{esbm2}, a BCH code with a length 16384 and $t=72$ is presented as an example to reveal the effectiveness of the proposed early-stopped checking method. For ES version 3 with $\kappa = 6$, the highest probability of undetected errors is calculated as $6.49437\times10^{-119}$ over the cross-over probability at $2.5\times10^{-3}$.
It can be shown as an example that ES version 3 provides a reliable result for early termination checking by observing that the number of discrepancies is consecutively zeros. For practical applications, the proposed ES version 3 should be considered to prevent decoding failure over the firmware and decoder commuting period. As a matter of fact, the reliability of the early stopped method is the major concern for the flash controller rather than comparing the performance. If the detection failure occurred from the BCH decoder, the credibility of hard decoding would collapse. To address this issue, this paper focuses on the practical consideration to investigate the malfunction probability in this sense. To evaluate the credibility of the proposed method, we have given a complete test sample based on an FPGA board from the Altera family Statix II which operates at a clock rate of 110Mhz and uses BCH code length of 16384 that is suitable for a USB firmware testing. The system throughput is set to 480Mbps based on the USB 2.0 standard. The whole test sample quantity has a great amount of $5.9793\times10^{35}$. Each test sample contains the data package of 3 BCH code blocks and the code length is 16383 using $GF(2^{14})$ BCH code t=72. This result means that we never encountered any decoding failure during the time using a storage device based on the proposed design. Finally, this technique has been applied to commercial USB devices since 2012 and the USB controller name is BR825CA illustrated in Fig.~\ref{825ca}. 
\section{Conclusion}
We have provided a practical solution for early termination checking while decoding BCH code. The complexity analysis and numerical results are presented to show the merit of the proposed technique which is suitable for long and large error correcting capability of BCH code with complexity reduction up to 80\% over conventional early-stopped approach in \cite{Liu}. The decoding failure is successful in exchange for decoding latency since the numerical result illustrates that the probability of undetected errors is lower than $6.49437\times10^{-119}$ for $GF(2^{14})$ BCH code t=72. The FPGA testing on a USB device using 16384 code length of BCH code has been implemented to justify the reliability of the early termination checking strategy and the number of testing samples is accumulated up to $5.9793\times10^{35}$. This approach is shown to provide a solution for a practical design.
\bibliographystyle{IEEEtran}
\bibliography{esbm010821}

\end{document}